\documentclass{iaus}
\usepackage{graphicx}

\title[Physical Properties Red Supergiants] 
{The Physical Properties of Red Supergiants: Comparing Theory and Observations}

\author[Philip Massey et al.]   
{Philip Massey$^1$, Emily M. Levesque$^2$, \\ Bertrand Plez$^3$, \and K. A. G. Olsen$^4$}

\affiliation{$^1$Lowell Observatory,
\\ 1400 W. Mars Hill Rd., Flagstaff, AZ 86001, USA
\\ email: {\tt Phil.Massey@lowell.edu} \\[\affilskip]
$^2$Institute for Astronomy, University of Hawaii,
\\ 2680 Woodlawn Drive, Honolulu, HI 96822, USA
\\email: {\tt emsque@ifa.hawaii.edu } \\[\affilskip]
$^3$GRAAL, Universit\'{e} Montpellier II, CNRS, 34095 Montpellier, France
\\ email: {\tt Bertrand.Plez@graal.univ-montp.fr} \\[\affilskip]
$^4$Gemini Science Center, NOAO
\\P.O. Box 26732, Tucson, AZ 85726-6732, USA
\\ email: {\tt kolsen@noao.edu}
}

\pubyear{2008}
\volume{250}  
\pagerange{}
\jname{Massive Stars as Cosmic Engines}
\editors{F. Bresolin, P. A. Crowther, \& J. Puls,  eds.}
\begin{document}

\maketitle

\begin{abstract}
Red supergiants (RSGs) are an evolved stage in the life of intermediate massive stars ($<25M_\odot$). For many
years, their location in the H-R diagram was at variance with the evolutionary models. Using the MARCS
stellar atmospheres, we have determined new effective temperatures and bolometric luminosities for RSGs in
the Milky Way, LMC, and SMC, and our work has resulted in much better agreement with the evolutionary
models. We have also found evidence of significant visual extinction due to circumstellar dust. Although
in the Milky Way the RSGs contribute only a small fraction ($<1$\%) of the dust to the interstellar medium
(ISM), in starburst galaxies or galaxies at large look-back times, we expect that RSGs may be the main dust
source. We are in the process of extending this work now to RSGs of higher and lower metallicities using
the galaxies M31 and WLM.
\end{abstract}

\firstsection 
\section{Introduction}

Those of us here who have worked on massive stars for a while are probably
all attracted by stellar physics at the extremes.  For O-type stars, we are dealing
with stars that are as massive and luminous as stars come, and as hot as main-sequence stars can get.  To properly assess their physically properties through
spectroscopic analysis has required not only the introduction of non-LTE
atmosphere models 
(\cite[Mihalas \& Auer 1970]{MihAue70},
 \cite[Auer \& Mihalas 1972]{AueMih72})
but an additional thirty years of developments, such as the inclusion of mass loss
(\cite[Abbott \& Hummer 1972]{AbbHum72},
 \cite[Kudritzki 1976]{Kud76}), the
inclusion of hydrodynamics of the stellar wind (\cite[Gabler, Gabler, Kudritzki, et al.\ 1989]{Gab89};
\cite[Kudritzki \& Hummer 1990]{KudHum90};
\cite [Puls, Kudritzki, Herrero, et al.\ 1996]{Puls96}), and,
most recently, the full inclusion of line blanketing (\cite [Hillier \& Miller 1989]{HilMil89};
\cite[Hillier, Lanz, Heap, et al.\ 2003]{Hil03}; \cite[Herrero, Puls, \& Najarro 2002]{Her02}). (For a recent
summary, see \cite[Massey, Bresolin, Kudritzki, et al.\ 2004]{OPapI}
and \cite[Massey, Puls, Pauldrach, et al.\ 2005]{OPapII}).  The study of LBVs and WRs is
equally exciting, stars where radiation pressure dances with gravity to see who 
will lead (\cite[Lamers 1997]{Lam97}, \cite[Smith \& Owocki 2006]{SmiOwo06}),
and where high mass loss rates are continuous rather than episodic due to 
high metal content in the stellar atmosphere (\cite[Crowther 2007 and references
therein]{Cro07}).

However, largely ignored until now are the
red supergiants (RSGs).  The physical conditions in these stars are, in their own way,
equally extreme.  They have the largest physical sizes of any stars (up to 1500$\times$
the radius of the sun; see \cite [Levesque, Massey, Olsen, et al.\ 2005, here after Paper I]
{RSGPapI}).  This large physical size invalidates the usual assumptions of plane parallel geometry.  The velocities of the convective layers in these stars' atmospheres are
supersonic, giving rise to shocks (\cite [Freytag, Steffen, \& Dorch 2002]{Fre02}), and making the stars'
photospheres very asymmetric and invalidating mixing-length assumptions.
Their extremely cool temperatures (3400 - 4300~K) lead to the the presence
of molecules in their atmospheres, requiring the inclusion of extensive 
molecular opacity sources in any realistic model atmosphere.  From an observational
point of view, the large (negative) bolometric corrections and their sensitivity to
the adopted temperature complicate the transformation from the observed color-magnitude diagram to the physical H-R diagram in much the same way as it does
for the O-type stars.

Recent advances in stellar atmosphere models for cool stars (e.g., \cite[Plez 2003]{Plez03})
have allowed the
first reasonable determination of the physical properties of these stars, in much
the same way that the  non-LTE H and He models of 
\cite{AueMih72} allowed the first reasonable determiation
of the physical properties of O-type stars by \cite{Con73}.  And while we may still
have a way to go, we believe our answers will hold up as well as those that
\cite{Con73} have, which is really pretty well (see \cite[Massey, Puls, Pauldrach, et al.\ 2005] {OPapII}).

I find it personally interesting that there has been a real aversion to looking at what
happens to a massive star as it heads over to the far right side of the H-R diagram.
I think this is cultural---for many years much of 
the ``massive star community"
really thought of itself as the ``hot star community", 
with the exception of a few workers,
most notably our good colleague Roberta Humphreys, whose early work on supergiants
in the Milky Way and other Local Group galaxies (such as
 \cite[Humphreys 1978] {Hump78},
 \cite[1979a]{Hump79a}, 
 \cite[1979b]{Hump79b},
 \cite[1980a]{Hump80a},
 \cite[1980b]{Hump80b},
 \cite[1980c]{Hump80c},
 \cite[Humphreys \& Davidson 1979]
 {HumpDavid79},
 \cite[Humphreys \& Sandage 1980]
 {HumpSand80})  
 certainly spurned my own interest
in the field, and whose presence at these  symposia always reminds
us that there's more to the life of a massive star than the O and WR stages. 

\begin{figure}
\begin{center}
 \includegraphics[width=2.65in]{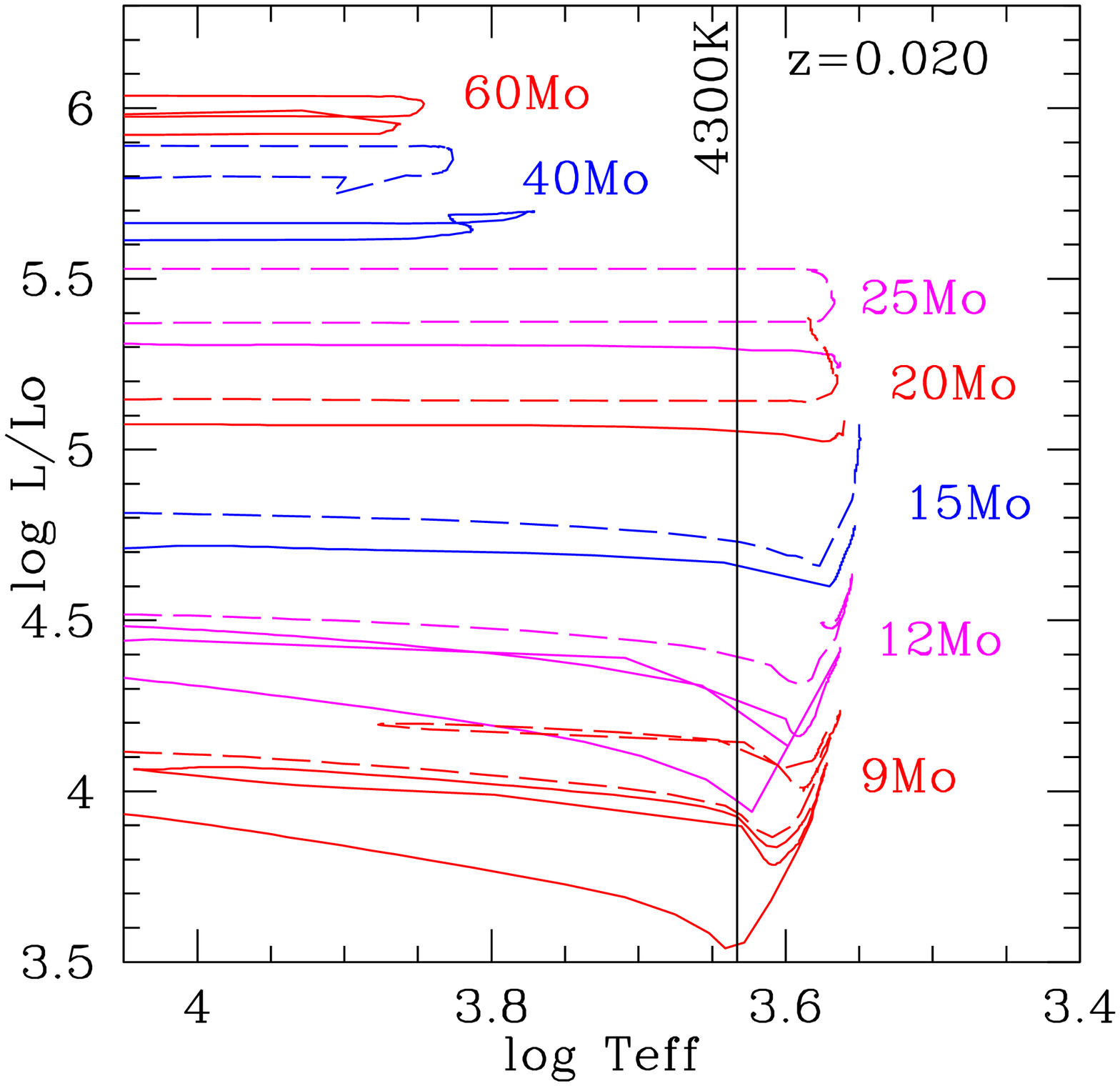} 
 \end{center}
 \includegraphics[width=2.65in]{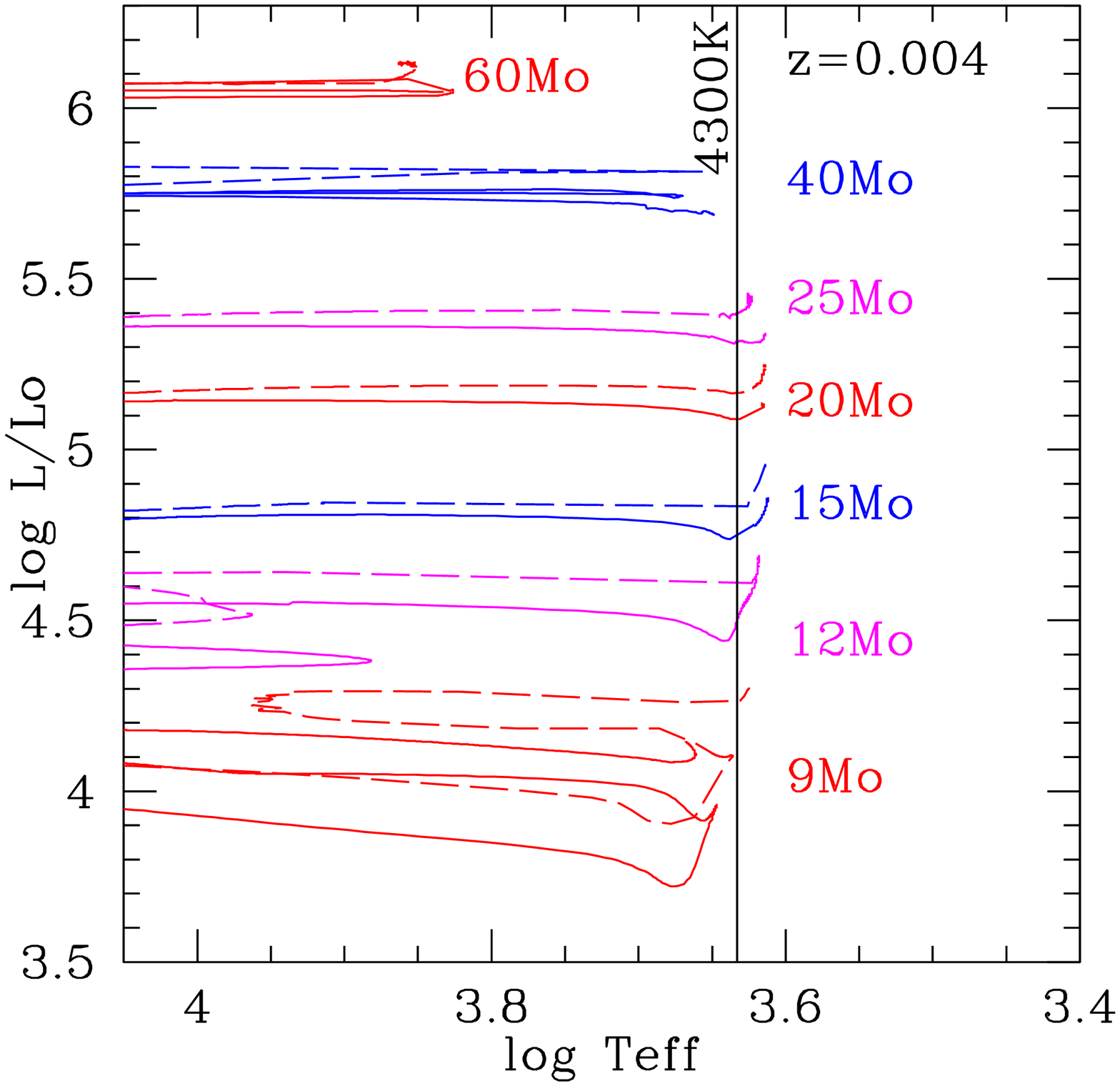} 
 \includegraphics[width=2.65in]{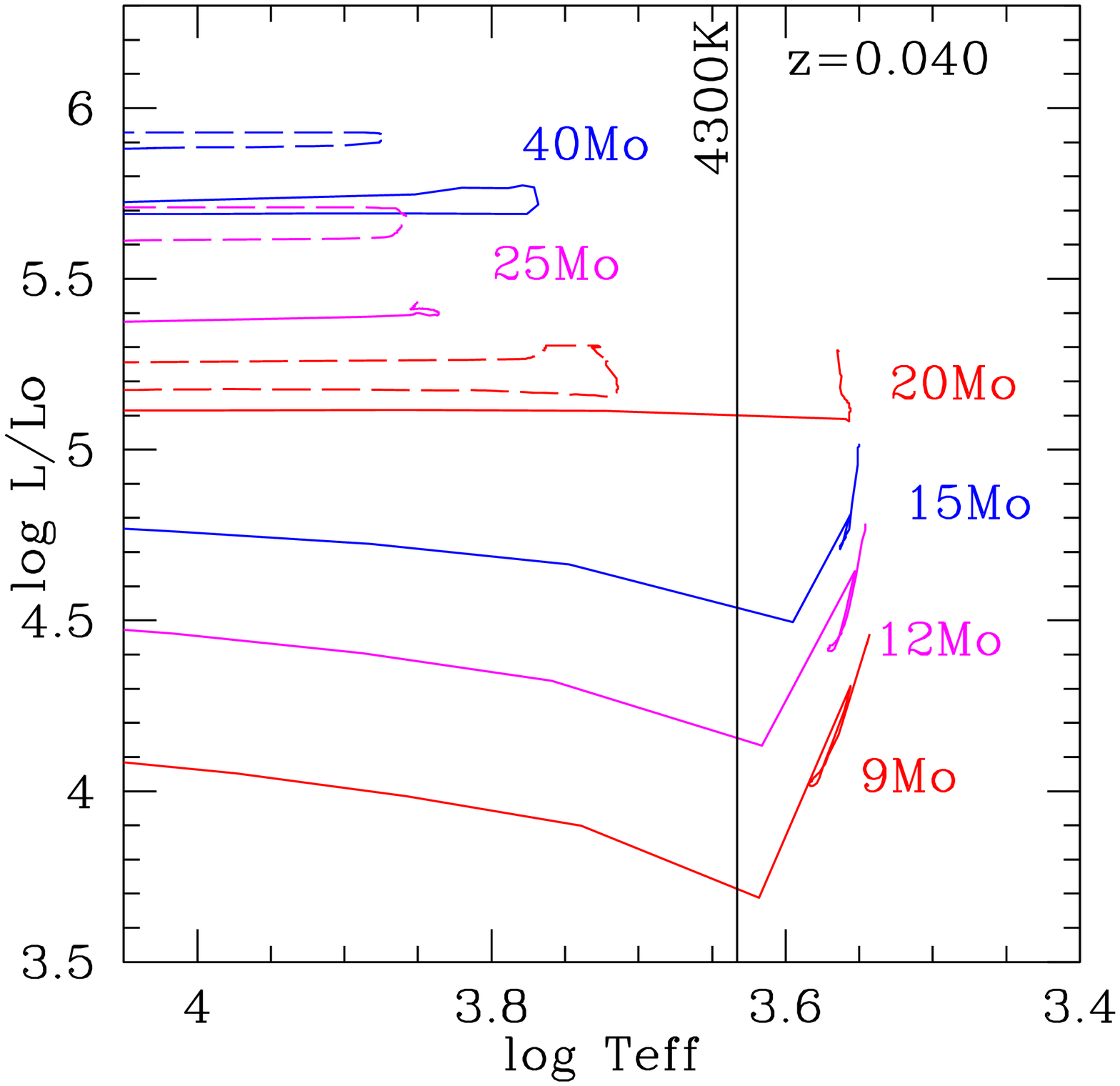}
 \caption{\label{fig:HRDs}Effects of metallicity on the evolutionary tracks in the RSG region.
The tracks for z=0.020
(solar) are from \cite{MM03}, for z=0.004 (SMC-like) are from \cite{MM00}, and for z=0.040 (M31-like) are from \cite{MM05} and \cite{MM94}. Solid curves denote the tracks with no initial rotation,
while the dashed lines correspond to initial rotations of 300 km s$^{-1}$.  The black
vertical line marks a temperature of 4300 K, roughly that of a K0~I at both solar
and SMC metallicity (Papers I, II).}
\end{figure}

Let us briefly review what the evolutionary tracks predict RSGs come from.
 In Fig.~\ref{fig:HRDs} we show the tracks covering a range of a factor of
 10 in metallicity, from z=0.004 (SMC-like) to z=0.020 (solar) to z=0.040 (M31-like).
 I have drawn in a vertical line at an effective
 temperature of 4300~K, which roughly corresponds to that of a K0~I. Stars to
 the right of that line we are calling RSGs.   At solar metallicities we expect that
 stars with initial masses $\leq 25M_\odot$ will become RSGs.  At lower metallicities
 (SMC-like) 
 the upper mass limit for RSGs is probably a bit higher---maybe $30M_\odot$?---it's
 hard to tell because of the quantization of the tracks. The upper
 mass tracks go much further to the right at this low metallicity, but stop short of
 the RSG dividing lines---these 30-60$M_\odot$ stars become F- and G-type
 supergiants, but not K or M.  At higher metallicity (M31) the
 limit is definitely lower, around $20M_\odot$.   In the case of solar metallicity the 
 $25M_\odot$ track
 turns back to the blue, and in fact such a star should become a WR after the RSG phase.
 
 One more thing of note is that the tracks don't extend very far to the right of the
 K0~I (vertical) line in the SMC---the RSGs in the SMC shouldn't be very late, mostly
 K through M0, say.  At higher metallicities they extend further to cooler effective temperatures.  This is consistent with the change in the average RSG type observed in
 the SMC, LMC, and Milky Way (\cite[Elias, Frogel, \& Humphreys 1985] {EFH85},
 \cite[Massey \& Olsen 2003]{MO03}).
 
That said, when we began worrying about the issue of the physical properties of RSGs
it was because if one placed the ``observed" location of RSGs on the H-R diagram,
they missed the tracks entirely: the alleged effective temperatures and luminosities
were cooler and higher than those predicted by the evolutionary tracks. This
was first noticed by \cite{MO03} for the SMC and LMC, but we quickly confirmed
that the problem also existed for the Milky Way sample (\cite[Massey 2003a]{MasARAA}).
When I mentioned this issue at the Lanzarote meeting (\cite [Massey 2003b]{MasIAU212}) Daniel Schaerer came up to me afterwards and said really, this was not a problem, since how far to
the right the tracks went were heavily dependent upon such issues as how the mixing length
was treated (see, for example, \cite[Maeder \& Meynet 1987]{MM07}).  However, this did not
explain the issue of the luminosities being too high, and as an observer
I was more concerned with what if the ``observations" were wrong?  Because, of course we don't
``observe" effective temperatures and bolometric luminosities; instead, we obtain photometry
and spectroscopy and use some relationship to convert these to physical properties.  Indeed,
further reading convinced me that there could be a serious problems, as much of what we 
``knew" about the effective temperature scale of RSGs were derived from lunar occultations of
red {\it giants} (not supergiants); see discussion in \cite{MO03}.

What would be involved in determining the effective temperatures of RSGs ``right"?  
We really need models that
have enough physics in them to correctly reproduce temperature-sensitive spectral features.
The participants at this conference (mostly) understand what was involved in getting there for
O-type stars.  RSGs present their own challenges, as noted above.  Fortunately, at the time I
got intrigued by this problem, sophisticated models that were up to the task were becoming available.  
A modern version of these MARCS models was described by \cite{Plez92}, based upon
the earlier work of \cite{Gust75}.   These are static, LTE, opacity-sampled
models, and the current version (\cite[Plez 2003]{Plez03}; \cite[Gustafsson, Edvardsson, Eriksson, et al.\ 2003]{Gust03}) includes improved atomic and molecular opacities and sphericity.

\section{RSGs in the Milky Way}

In Paper I we obtained moderate-resolution
spectrophotometry of 74 Galactic RSGs, which we then compared to the models.
Our primary selection criterion was that the RSG had to be in a cluster or an
association with a relatively well determined distance from the OB stars (\cite[Humphreys 1978]{Hump78}, \cite[Garmany \& Stencil 1992]{GS92}).
We used a grid of models with effective temperatures 3000-4500~K in increments
of 25~K, and $\log g = -1$ to +1 [cgs] in steps of 0.5~deg.   We would typically begin
by reddening the $\log g=0.0$ model spectra of various effective temperatures by
different amounts (using a \cite[Cardeli, Clayton, \& Mathis 1989]{CCM89}
 reddening law with $R_V=3.1$) until we got a good match to the depths of the 
 molecular bands (principally TiO) {\it and} continuum shape of the spectra of the star.
 We would then see if the derived luminosity implied a surface gravity consistent
 with the $\log g=0.0$; if not, we used a more appropriate model.  In practice, the
 value we determined for the effective temperatures did not depend upon our
 0.5~dex uncertainty in the adopted surface gravity, and the $A_V$ was affected by $<0.3$~mag.  
  
\begin{figure}[h]
 \begin{center}
 \includegraphics[width=3.5in]{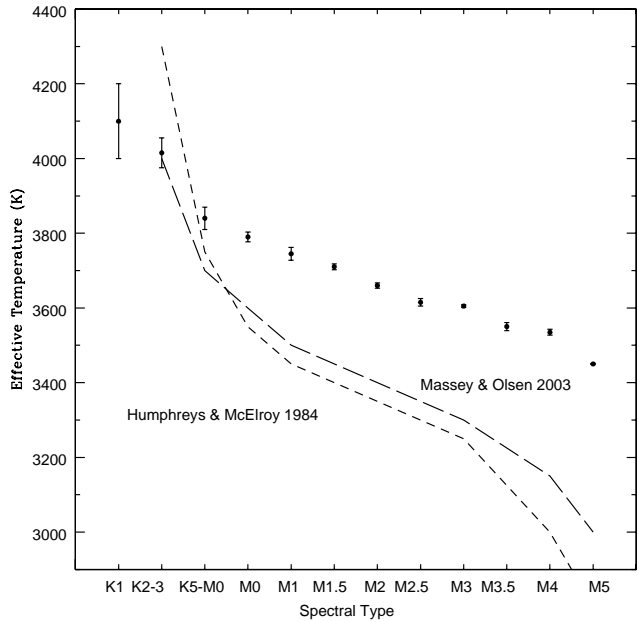} 
 \caption{\label{fig:MWscale} Effective temperature scale for Galactic RSGs.  The
 new temperature scale for RSGs is shown by the points.  For comparison, we include the much cooler effective temperatures of \cite{HumpMc84} and \cite{MO03}.}
\end{center}
\end{figure}

 When all was said and done, we had derived a new effective temperature scale
 which was significantly warmer than the older ones.  It is shown by the points
 in Fig.~\ref{fig:MWscale}.  Their error bars reflect the standard deviation of the
 mean of our determinations; for the M stars (where we can use the TiO bands) our
 precision was $\sim$25~K, or about 0.7\%---compare this to the typical
 2000~K (5\%) uncertainty we have when fitting O stars and their weak lines!
 
\begin{figure}[h]
 \includegraphics[width=2.65in]{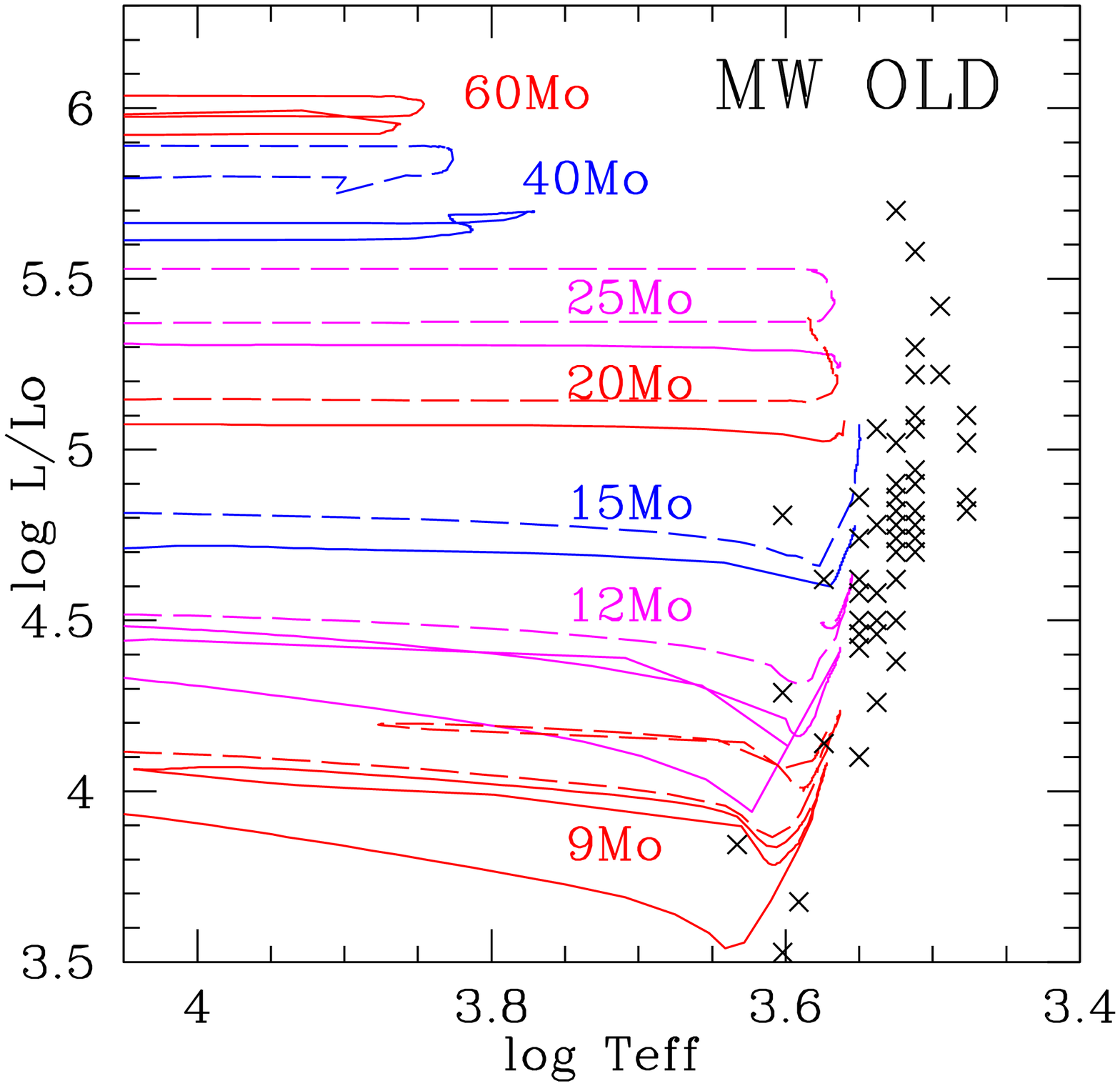} 
 \includegraphics[width=2.65in]{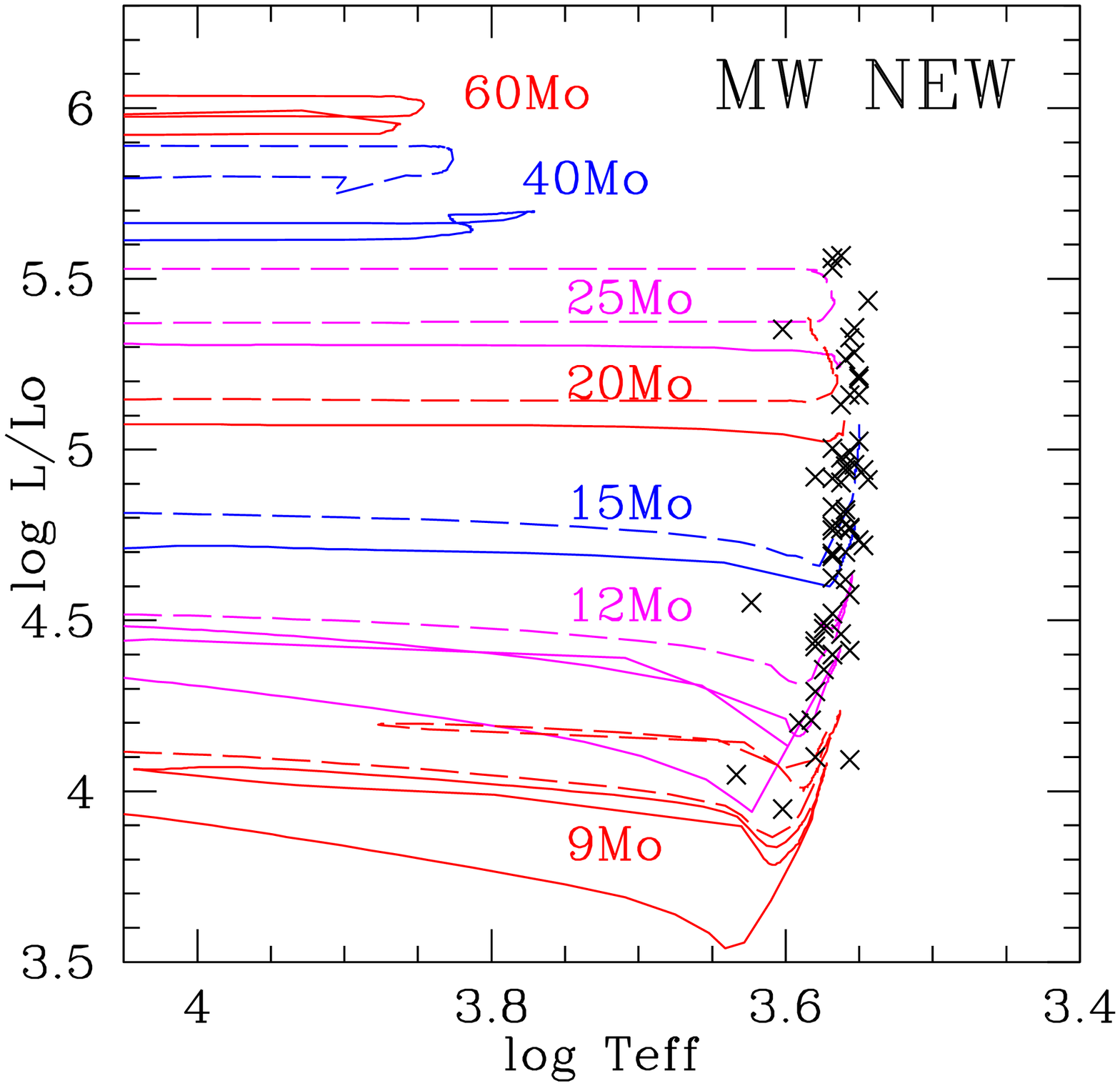}
 \begin{center}
  \includegraphics[width=2.65in]{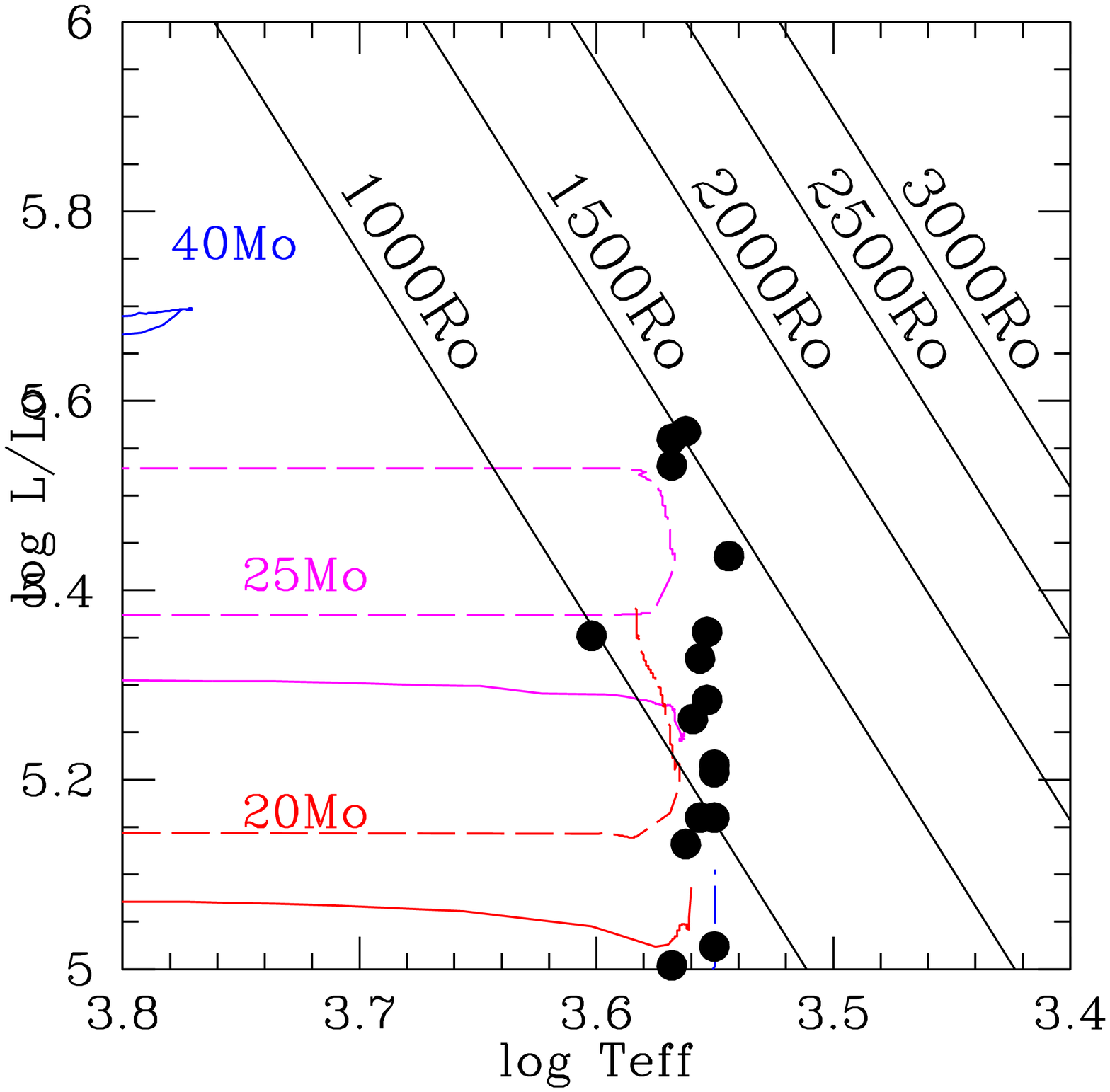}
  \end{center}
 \caption{\label{fig:MWHRDs} Agreement with evolutionary tracks for the Milky Way.  On
 the left (top) we show the agreement (or rather lack thereof) between the
 evolutionary tracks and  the ``observed"
 location of RSGs using the old effective temperature scale and bolometric
 corrections given by \cite{HumpMc84}.  On the right (top), we show the agreement
 using the results from Paper I.  The evolutionary tracks are the same as those
 shown for $z=0.020$ in Fig.~\ref{fig:HRDs}  In the bottom figure we show an
 expansion of the upper-right part of the later figure, now with lines of constant
 radii indicated.}
\end{figure}

What did that do to the placement
 of stars in the H-R diagram?  Just what we hoped!  We show the situation (old and new)
 in Fig.~\ref{fig:MWHRDs}.  Now there is excellent agreement both in the effective
 temperatures, and in the upper luminosities, of RSGs in the Milky Way.

One of the cute things to come out of Paper~I is the answer to ``How large do normal stars get?"
We see at the bottom of Fig.~\ref{fig:MWHRDs} a blowup of the upper right of our H-R diagram,
now with lines of constant radii marked.    The largest stars known in the Milky Way have radii of
1500$R_\odot$, or about 7.2~AU.  If you were to take one of these behemoths, and plunk it down
where the sun is, its surface would extend to between the orbits of Jupiter and Saturn.  Of course,
``real" RSGs are known to have highly asymmetrical and messy ``surfaces", as witness the high
angular resolution images obtained of Betelgeuse by \cite{Young00}.  

\section{RSGs in the Magellanic Clouds: Weirder and Weirder}

\begin{figure}[h]
 \includegraphics[width=2.65in]{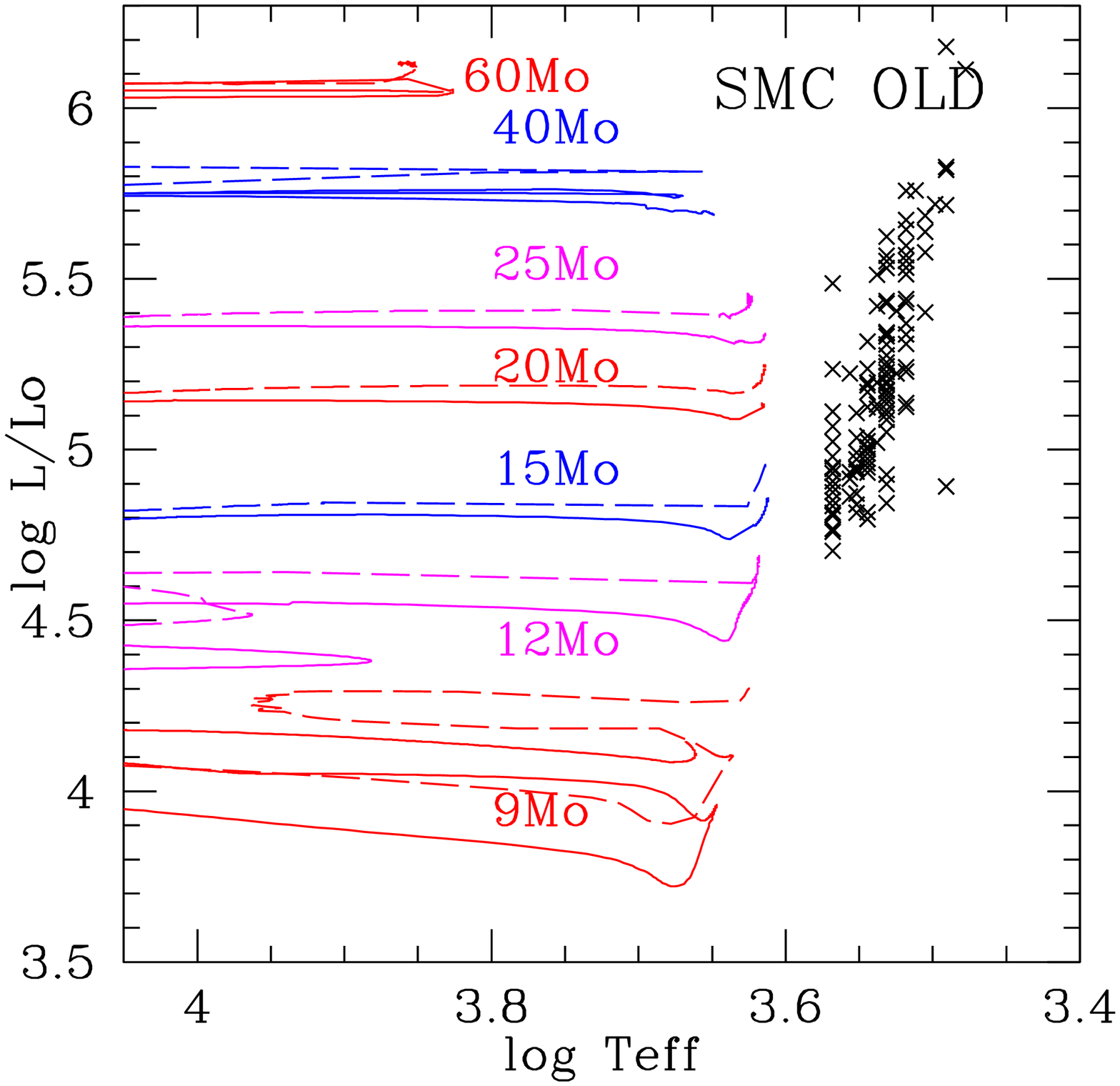} 
 \includegraphics[width=2.65in]{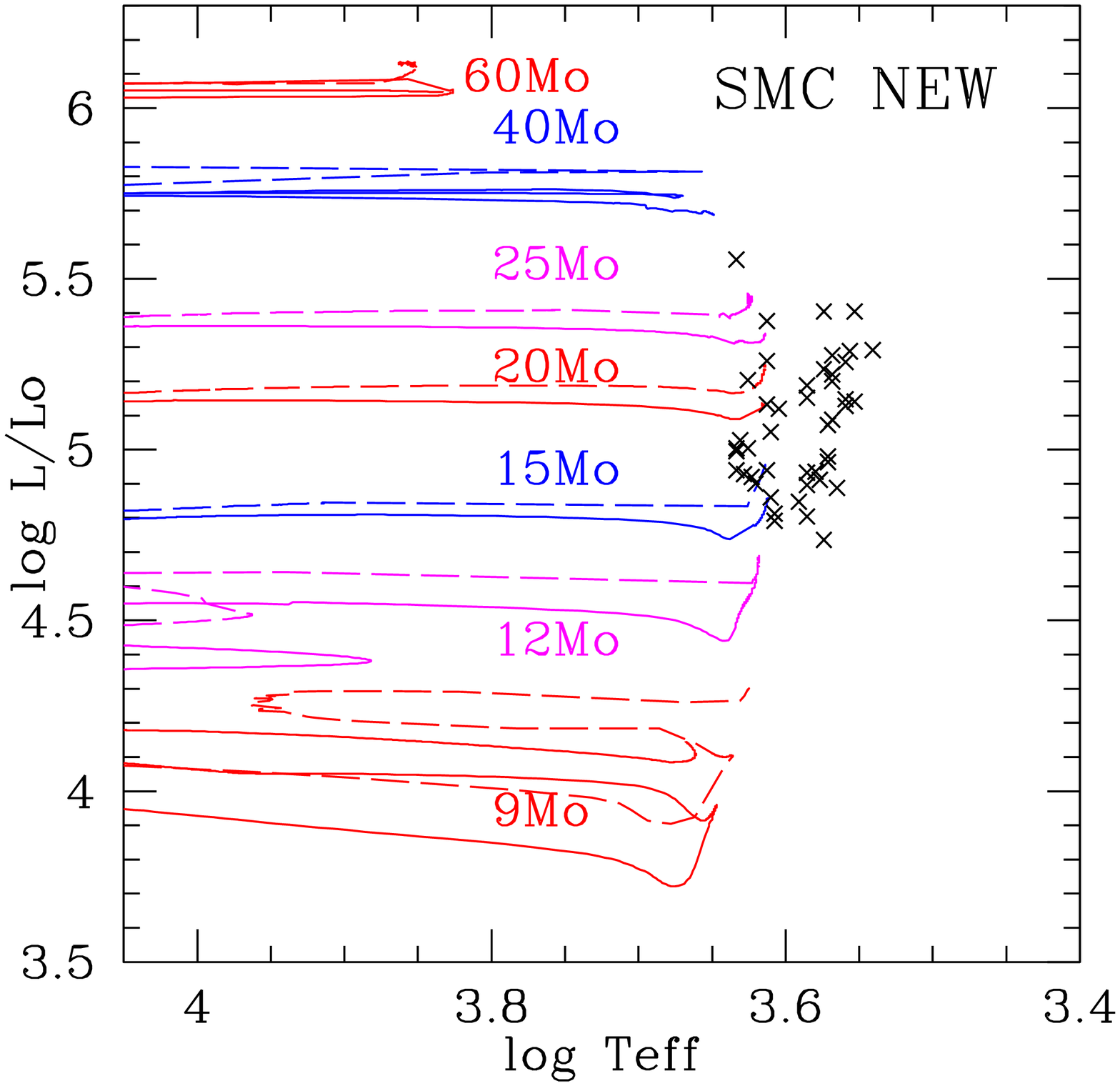}
  \caption{\label{fig:HRDSMC} Agreement with evolutionary tracks for the SMC.  On
 the left we show the lack there of)between the
 evolutionary tracks and  the ``observed"
 location of RSGs.  On the right, we show the agreement
 using the results from Paper II.  The evolutionary tracks are the same as those
 shown for $z=0.004$ in Fig.~\ref{fig:HRDs}.}
 \end{figure}

We naturally wanted to extend this work to RSGs in the Magellanic Clouds, where the metallicities
are lower than in the Milky Way.  Since the metallicity is lower, we expect that we will need cooler
temperatures in order to form the same strength of TiO, the basis for the classification of mid-K through
M stars.  And, indeed that's just what we found (Paper II): M-type stars are 50 K and 150 K cooler in
the LMC and SMC, respectively, compared to their counterparts in the Milky Way.
 Just as we had for Galactic RSGs,  we found
great improvement between the observations and the models.  For the LMC there is excellent agreement (not shown here; see Fig.~8 in Paper II).  For the SMC the results were also a great
improvement (Fig.~\ref{fig:HRDSMC}),  but there were a substantial number of stars that were a
bit cooler than the tracks allow.

This ``no star zone" beyond the end of the tracks is known as the Hayashi forbidden region---stars
in this region are no longer in hydrostatic equilibrium.   They shouldn't exist.   Even before we had
these results, we were intrigued by the fact that there were {\it some} stars in the LMC and SMC that
were classified as significantly later than the average type by \cite{MO03}.  

However, the real
revelation came in our efforts to obtain a spectrum of HV~11423, one of the brightest RSGs in the
SMC.  It was on our observing list because it was classified as an M0~I by \cite{EFH85} (based
upon photographic spectra obtained in 1978 and 1979), and we were tired of all of the K-types
we had been observing.  But, when we took spectra of it in early December 2004 it appeared
to be of early K-type, probably K0-1~I.  We honestly didn't think much about this at the time, but
imagined that perhaps we had gotten the wrong star, although the two spectra we had obtained
(on different nights) had matched.  The next year  (December 2005) we tried again, and took
a couple of spectra.  Much to our amazement the star was M4~I, much, {\it much} later than the
spectral types seen for RSGs in the low metallicity SMC.  We then went back and checked, and
of course we {\it had} taken the spectrum of the correct star in 2004 as well---the coordinates left
no doubt of that.  We took another spectrum the following year, in September (2006), and the
star was again an early K.  Some digging in various data archives unearthed a VLT/UVES spectrum
(apparently never published) taken in December 2001; the star was clearly of even later type
than our December 2005 M4~I type---more like an M4.5-5~I.  So, here is one of the {\it brightest}
RSGs in the SMC, and it is doing this funny little jig in the H-R diagram, changing effective temperatures
from 3300 to 4300 K on the time scale of months and no one had noticed.  Furthermore, the amount
of visual extinction ($A_V$) changed by more than
1~mag, which we attribute to episodic dust formation (see below).  
Details can be found in \cite{HV11423}.

We concluded that this star is in an unstable period, maybe near the end of its life.   Of course, one
star is an oddity.  The wonderful thing is that \cite{toocool} found several more just like it!  We think this
underscores just exactly how lightly we've scratched the surface of stellar population studies of even the nearest galaxies.

\section{Self-Consistency:  Broad-band colors and VY CMa}

\begin{figure}[t]
 \includegraphics[width=2.65in]{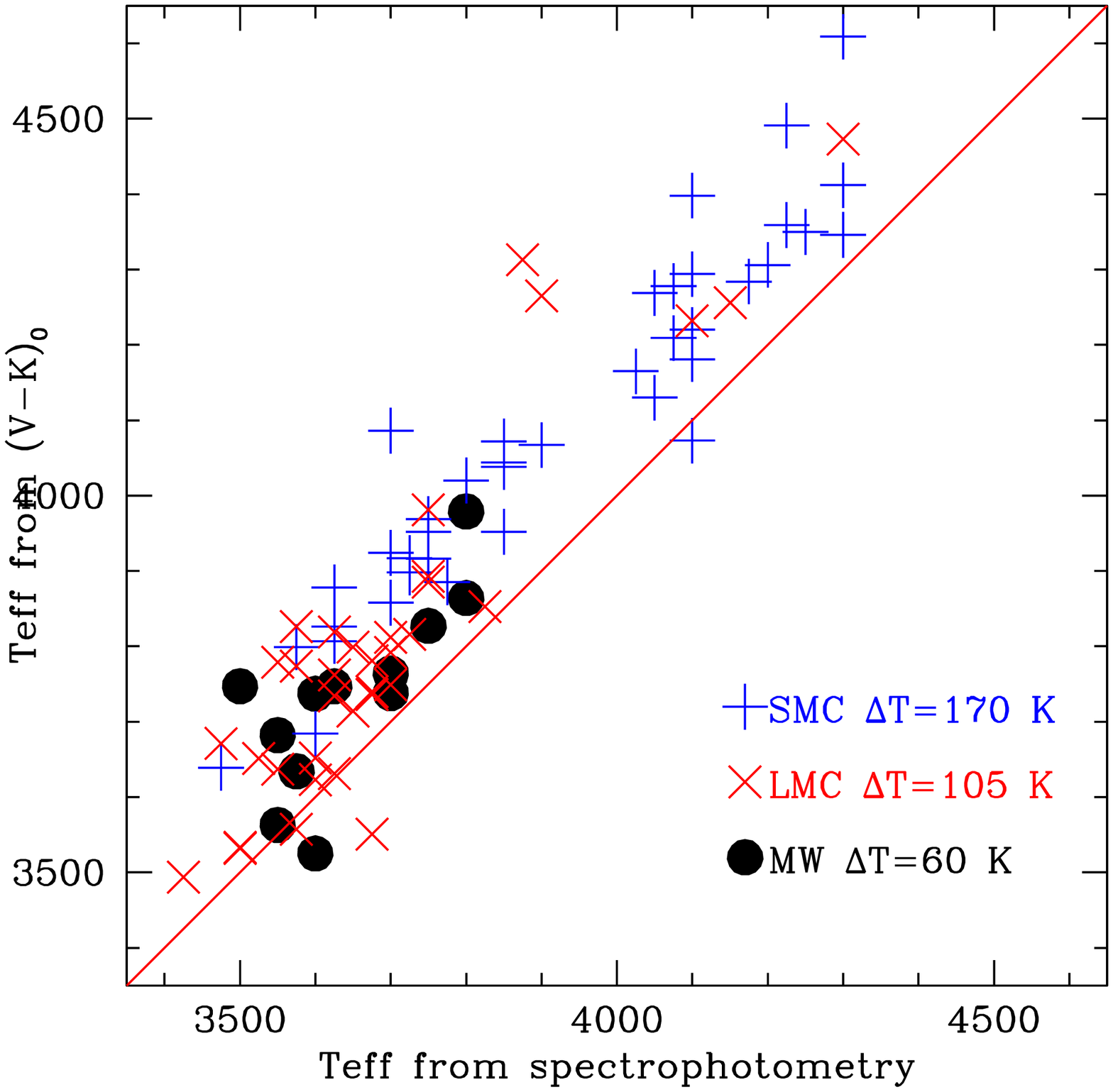} 
 \includegraphics[width=2.65in]{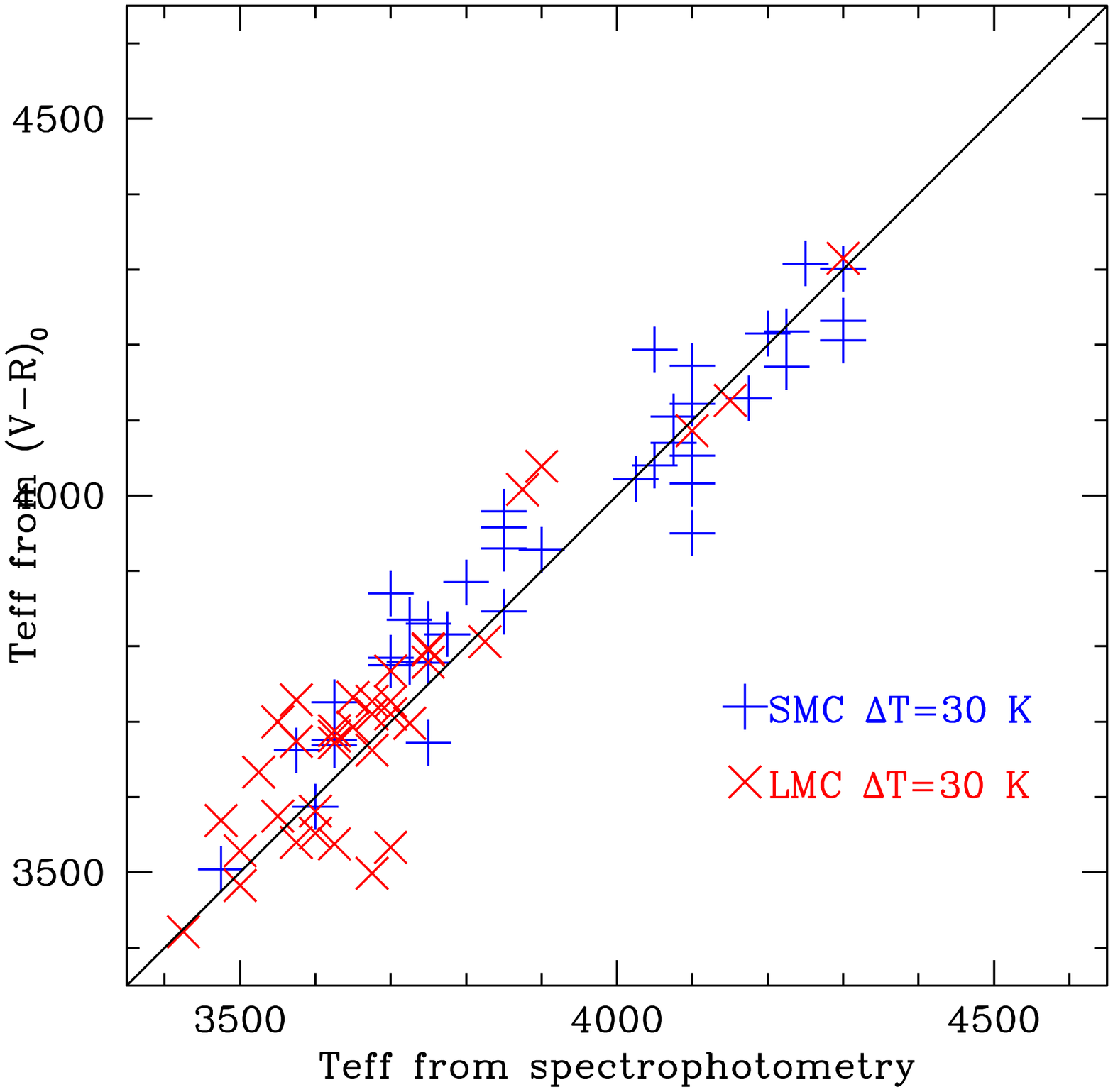}
  \caption{\label{fig:bb} The effective temperatures derived from broad-band photometry are compared
  to those determined from spectral fitting.}
 \end{figure}

If we only talked about our successes, we would be doing public relations and not science.  One of
the critical tests we performed was to see if our spectral fitting gave results that were consistent with
what we would get from the models if we were to use the broad-band colors $(V-K)_0$ and $(V-R)_0$ instead.  Such a test is  not completely independent from our spectral fitting, as we must deredden the 
broad-band colors to make these comparisons, and for this we adopt the reddenings determined
from the spectral fittings, but that is relatively minor.  In Fig.~\ref{fig:bb} (left)
we show the comparison
between the derived effective temperatures from the spectral fittings, and that obtained from 
the $(V-K)_0$ colors.  We see there is a systematic difference that is apparently
metallicity-dependent: the median difference is 60~K for the Milky Way, 105~K for the LMC,
and 170 K for the SMC, all in the sense the the effective temperatures derived from $(V-K)_0$
are hotter.  To make the SMC data conform we would have to finagle the $(V-K)_0$ calibration 
by nearly 0.5~mag, so this is not due to some sort of subtle photometric transformation issue from
$K_s$ to $K$ or something.  Our first thought was that this was some sort of discrepancy having to
do between the fluxes derived from the models and the strengths of the spectral features---the
$(V-K)_0$ effective temperatures depend upon the former, while the spectral fitting effective
temperatures depend upon the latter---but this notion was dispelled by looking at the results
from $(V-R)_0$.  Here we find very good agreement between the effectives temperatures
derived from spectral fitting and those from photometry.

 Instead, we now believe this is a discrepancy between the effective temperatures derived from
the optical and those derived from the near-IR, and may just be due to the intrinsic limitations of
static, 1-D models.  We know that these stars likely contain cool and warm regions on their
surfaces (\cite[Freytag, Steffen, \& Dorch 2002]{Fre02}), so it would not be unreasonable if the
effective temperature one measures is wavelength dependent.   This issue is discussed in greater
depth in Paper II.

Let us now turn briefly to an analysis we did of VY CMa (\cite[Massey, Levesque, \& Plez 2006]{VYCMa}), where we got both the right and wrong answers.  VY CMa is a Galactic RSG with
some extreme properties claimed for it in the literature: a luminosity of $2$ to $5\times 10^5 L_\odot$,
a mass-loss rate of $2\times 10^4 M_\odot$ yr$^{-1}$, with an effective temperature usually quoted
as 2800-3000~K.  We were disturbed by these values, as if you plotted the star in the H-R diagram
based on these, it would lie well into the Hayashi zone.  Yet, the star is fairly stable---George Wallerstein
has been observing it spectroscopically for many decades, and the only changes observed have to
do with weak emission that originates in the extensive nebulosity around the star.  We analyzed the
star based upon new optical spectrophotometry and existing optical and {\it JHK} photometry, and
concluded that the star had an effective temperature of 3650 K and a luminosity of $0.6 \times
10^5L_\odot$.

There was only one itsy-bitsy problem with these results: they had to be wrong. Once our paper appeared, several colleagues called our attention to the fact that the luminosity we derived
for the star was inconsistent with the total luminosity of the system (star plus dust).  We should have
realized there was a problem ourselves, as we had derived an effective temperature and
radius for the surrounding dust shell. The corresponding luminosity of the
dust  (which we did not work out) is 2.3$\times 10^5 L_\odot$,
 about $4\times$ larger than what we got for the star itself.
Since the dust is heated by the star, this is impossible.  

About the only way we have found out of this would be if there was substantial extra grey extinction. 
We get $A_V$ by reddening the models of appropriate 
temperature to match the shape of the stellar continuum using a $R_V=3.1$ \cite{CCM89} law.  However,
if the copious dust surrounding VY CMa has a distribution of grain sizes which is skewed towards
larger sizes than usual, then we would underestimate $A_V$ as the
dust would be greyer than we assume.
  We would need about an additional 1.5-2.0 mag of such grey extinction.  In any event, if we take our
  effective temperature, and a luminosity of 3 or 4$\times 10^5 L_\odot$ then the star sits in a very
  reasonable place in the H-R diagram, near the upper luminosity limit for RSGs.
  
  Still, we don't think this reveals some fundamental flaw with what we're doing.  The amount of dust
  around VY CMa is quite unusual  (see
  \cite[Smith, Humphreys, Davidson, et al.\ 2001]{SmithVY}), and it will be of interest to determine
  the properties of this dust.
  
\section{When Smoke Gets in Your Eyes}

\begin{figure}[b]
\begin{center}
\includegraphics[width=2.5in]{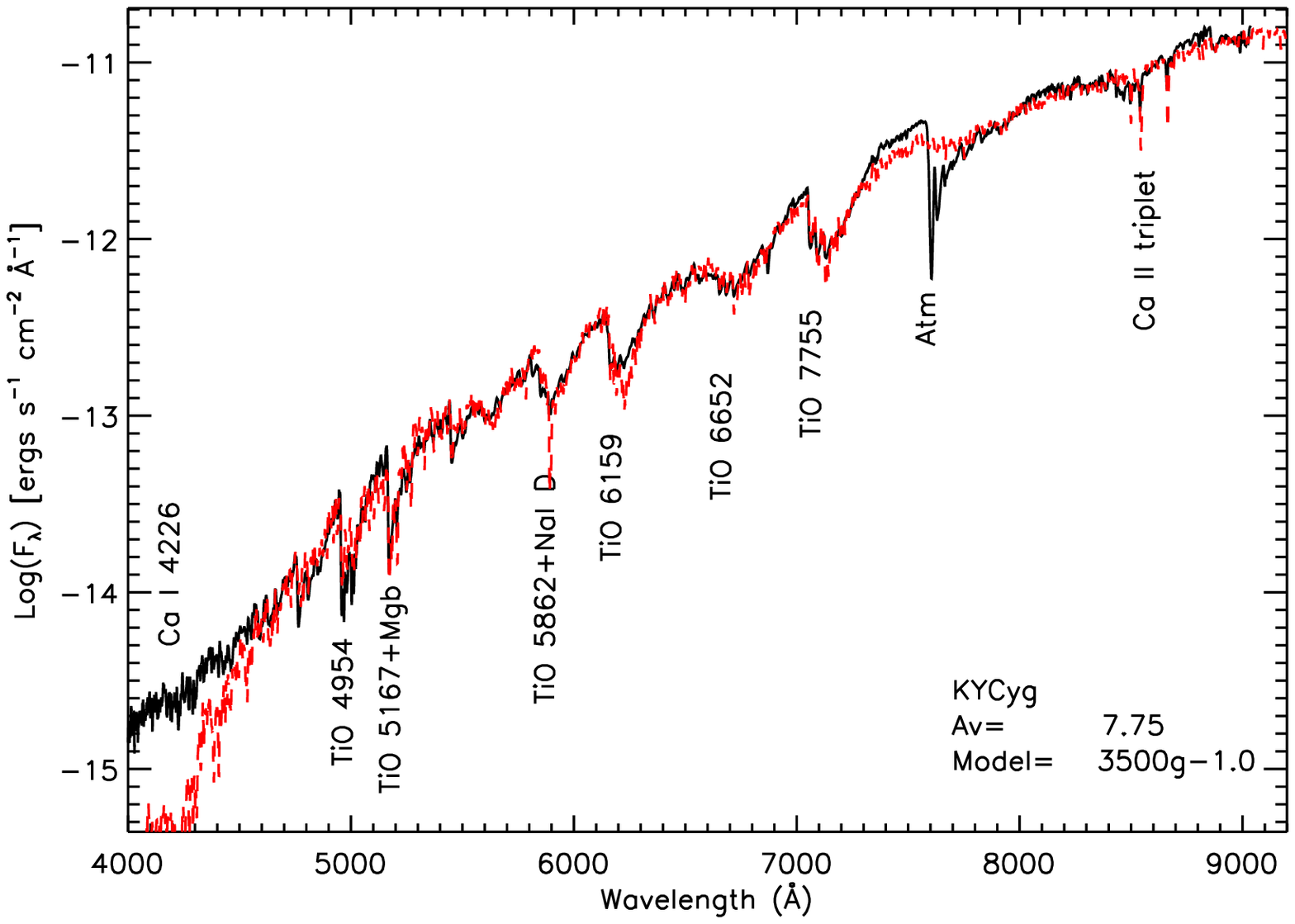} 
 \includegraphics[width=2.65in]{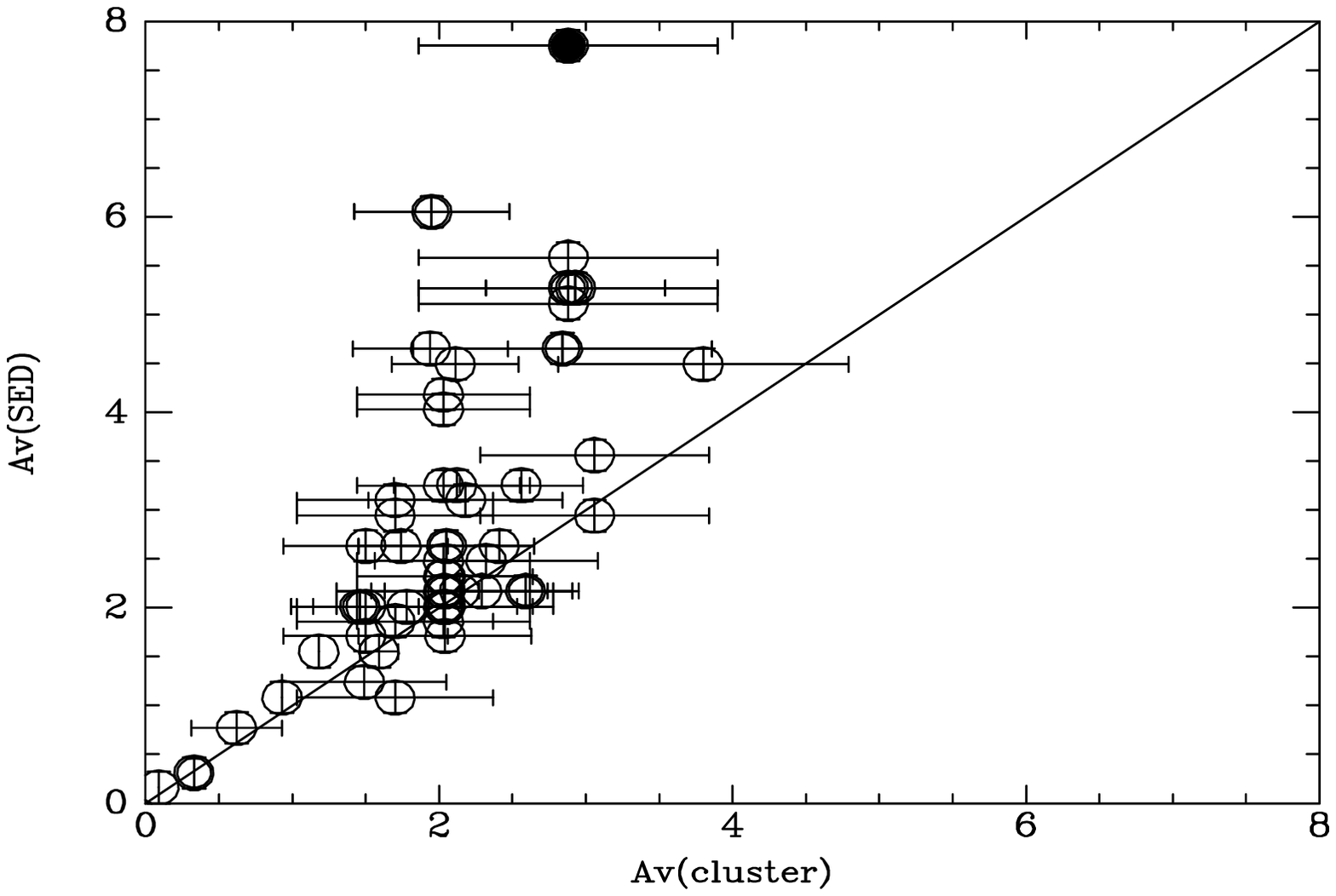} 
 \end{center}
  \caption{\label{fig:Av} The effects of circumstellar dust?
  On the left we show the NUV problem for the star KY Cyg.  Note that the star (solid line)
  has far more flux in the NUV than does
  the reddened model (dashed).  On the right the RSG extinction is 
  plotted against that of the OB stars in the same associations.  The error bars denote
  the range of extinction of the OB stars. The filled circle at the top denotes KY Cyg, which has one of the worst NUV problems.}
 \end{figure}

One of the things that worried us when we were doing our fits was that there were some stars
for which there was very poor agreement in the near-UV (i.e., $<$4100\AA, hereafter NUV), 
always in the sense
that the star had more flux than the best-fitting model.   
We illustrate an example in Fig.~\ref{fig:Av} (left) for
the star KY Cyg.
Now, we considered a number of possibilities.
Were these binaries, with the NUV being contributed by a hot companion?  We didn't think so.
We had indeed found some stars that clearly {\it were} binaries---but this was evident by having
a composite spectrum, with Balmer lines clearly evident.  The remaining stars that showed extra
flux in the near-UV didn't exhibit any signs of a composite spectrum. So that didn't wash as
an explanation.  

We investigated this further in \cite{smoke}. Was this due to a problem with the models?
We didn't think so, because there were lots of stars
that didn't show this problem, and there didn't seem to be any correlation with effective temperature.
What the NUV problem did show a correlation with was the amount of visual extinction---stars
with the largest NUV problem also had the largest $A_V$.  We looked into this a little more deeply, and indeed it
turned out that the stars with the largest $A_V$ actually had a considerable amount of {\it excess}
extinction compared to OB stars in the same OB associations.  
This is illustrated in Fig.~\ref{fig:Av}.
The error bars show the typical sigma of $A_V$ of the OB stars in a given association.  For most
RSGs there was good correlation between the $A_V$ found for the OB stars, and the $A_V$ found for
the RSGs, but for some significant fraction of the RSGs there was significantly more extinction---up to
several magnitudes\footnote{The alert reader may realize that $A_V$ determined by $E(B-V)$ from
broad-band colors requires a different effective $R'_V$ for intrinsically red stars than for intrinsically
blue stars.  We derive our
$A_V$ from spectrophotometry, so this issue doesn't apply.  Still, if you are trying to do this for
RSGs from broad-band photometry, then $R'_V=4.1+0.1E(B-V)-0.2\log g$; see discussion in
\cite{smoke}.} We see the same thing for RSGs in the Magellanic Clouds: in both the SMC and LMC
RSGs show greater extinction than the OB stars (Paper II).  

If the extra extinction was due to circumstellar dust, then that could also explain the extra flux in the
NUV---light near the star would be scattered by the dust, making the light more blue.
(This is different than the effect of dust along the line of sight,)   But, no one
had suggested that the observed
dust mass-loss rates would lead to significant circumstellar extinction around
RSGs.  We did the math, though (\cite[Massey, Plez, Levesque, et al.\ 2005]{smoke}), and this really
was {\it exactly} what you would expect: a thin-shell approximation (10 yr of dust mass loss at a
rate of $10^{-8} M_\odot$ yr$^{-1}$ condensing out at
distance of ten stellar radii) should lead to $>1$~mag of visual extinction.

Exactly how much dust do RSGs contribute
to the ISM?   In our work, we found a nice correlation between the bolometric luminosity of a star
and its dust mass-loss rate.  With this we were then able to estimate the amount of dust that RSGs
contribute locally, about $3\times10^{-8} M_\odot$ yr$^{-1}$ kpc$^{-2}$, in good agreement with the
value estimated by \cite{JurKle90}.   This is probably $3\times$ less than that contributed by late-type
WCs in the solar neighborhood, and about 200$\times$ less than that contributed by asymptotic giant
branch (AGB) stars.  So, locally, they don't amount to much as dust producers.  However, in a metal-poor
starburst, or in galaxies at large look-back time, one would expect  RSGs  to {\it dominate} the 
production of dust, as late-type WCs are not found in metal-poor systems, and AGBs require several
Gyr to form. 

Before leaving this subject, I'd like to briefly address the subject of RSG mass-loss.  We don't
know what drives the mass-loss of RSGs: , arguments have been presented both for pulsation and for
having the dust drive the wind.
But, I'd like to quote something my colleague Stan Owocki wrote in an email about all this, contrasting RSG mass-loss with O star mass-loss.    The escape velocity from a star is just 620 km s$^{-1} \times \sqrt{(M/R)}$.
O stars have a M/R ratio that is of order unity, but not RSGs!  There the ratio is much smaller,  more like 0.02.
So, the escape velocity is down by a factor of 7 or so, under 100 km s$^{-1}$.
Stan argued that the mass-loss of a hot star is set by conditions outside the stellar interior,
i.e., opacity in the atmosphere and wind, that results in the classic CAK mass loss (\cite[Castor, Abbott,
\& Klein 1975]{CAK75}).  A RSG, on the other hand, suffers mass loss because the ``heavy lifting"
has been done by the interior, as a significant fraction of the luminosity of the star has gone into
making a bigger radius.   So,
Stan argues, it is kind of like walking with a nearly full glass of water (RSG) vs a
glass that is only 1\% full (O star)---even a small jiggle can lead to big changes in the
mass loss for a RSG.

 \section{The Future}
 
Where do we go from here?  Our group is working on several projects.  One of this is to extend these
studies to other metal-poor galaxies, particularly WLM, and see if (for instance) we can find more
wacky late-type RSGs like HV~11423 and its friends.  Another is to extend this to M31, where the
metallicity is $2\times$ solar, at least according to studies of nebular abundances (for instance,
\cite[Zaritsky, Kennicutt, \& Huchra 1994]{ZKH}).  This brings us to one of our preliminary results.
Abundance studies of several M31 A- and B-type supergiants by 
\cite{Venn00} and \cite{Smartt01} found abundances that were essentially solar, not $2\times$ solar.
This is of course confusing, as it flies in the face of everything we know (or thought we knew!) about
one of our nearest neighboring galaxies.  

\begin{figure}[t]
 \includegraphics[width=2.65in]{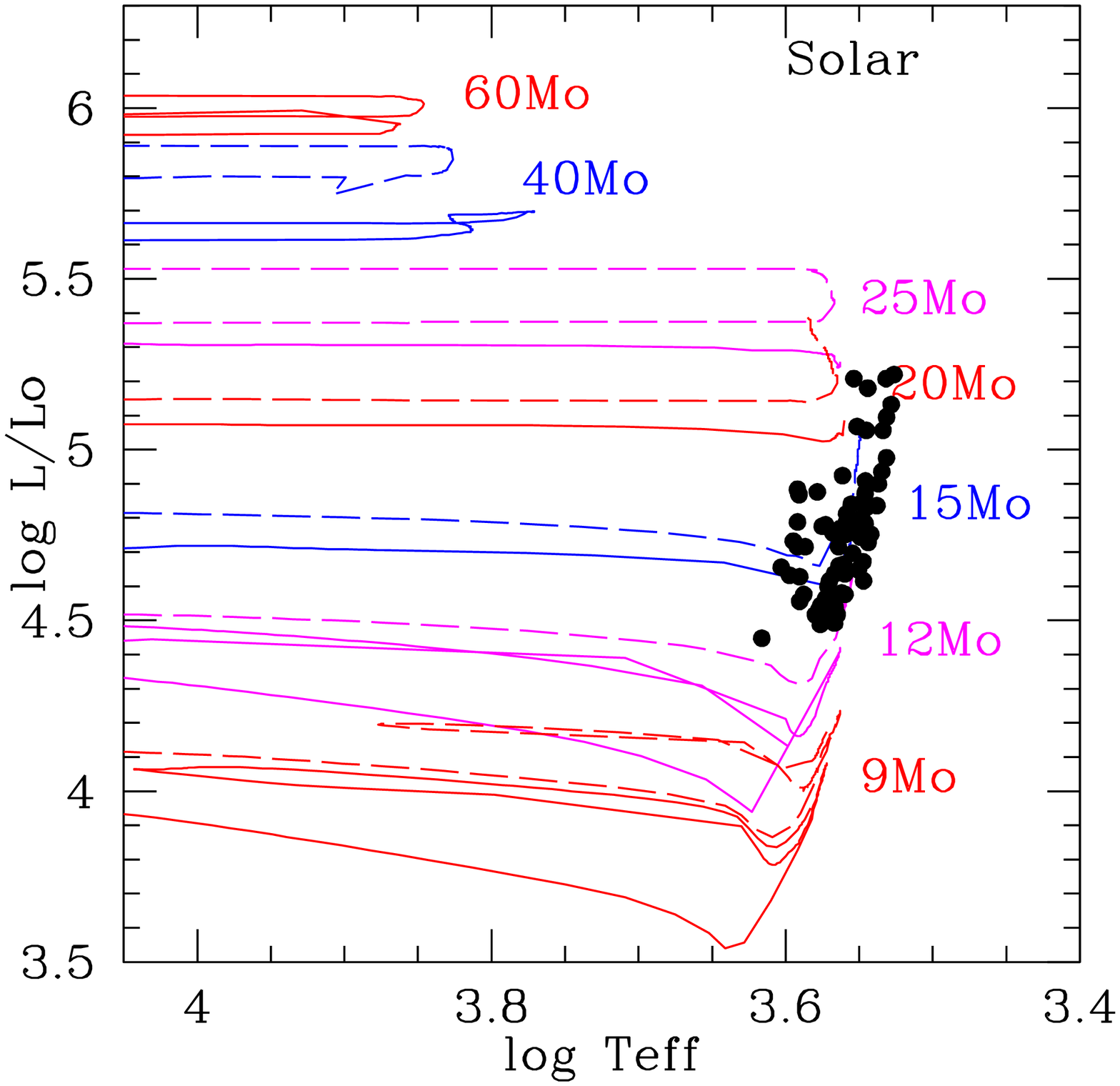} 
 \includegraphics[width=2.65in]{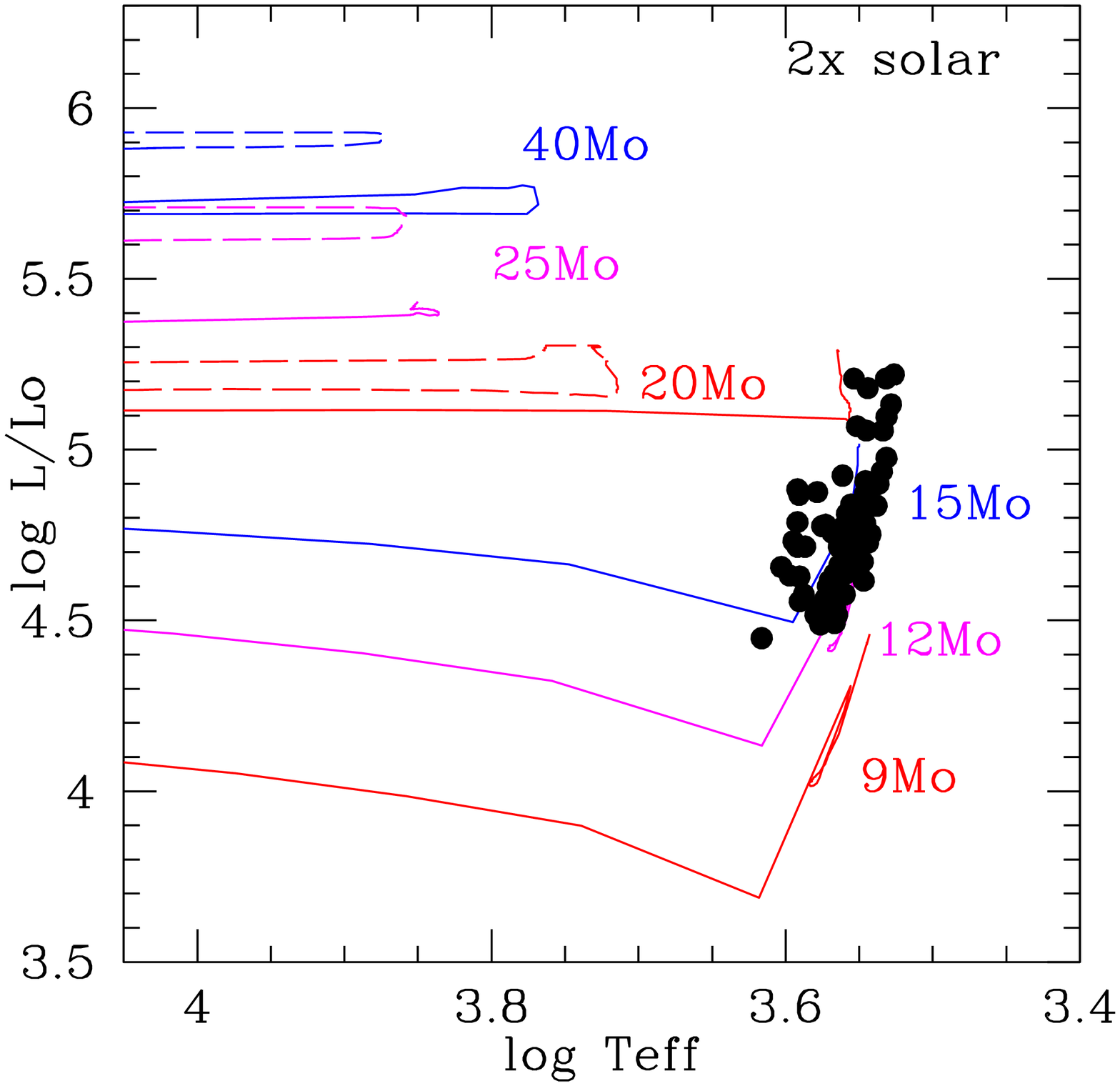}
  \caption{\label{fig:HRDM31} RSGs in M31 compared to solar (left) and $2\times$ solar (right)
  tracks.  The evolutionary tracks used are the same as in Fig.~1.  The distribution works well for the
  $2\times$ tracks, but not the solar tracks, which predict higher luminosity RSGs than what are
  observed.
   }
 \end{figure}

We can comment on this briefly.  The observed upper luminosity of the RSGs is consistent with that
expected based on the  $2\times$ solar
tracks but not the solar metallicity tracks. This is illustrated in Fig.~\ref{fig:HRDM31}.  If the metallicity were truly solar, then
where are of the high luminosity RSGs?  The ones that would come from $25M_\odot$?
(Compare also to Fig.~\ref{fig:MWHRDs}, upper right.)
But the 2$\times$ solar models work very well.  I was gratified to learn
at this conference that Norbert Przybilla finds similarly high abundances for these A-type
supergiants in his reanalysis, using  the
improved photometry that our Local Group Galaxies Survey has provided.

We thank our colleagues Georges Meynet and Andre Maeder, who co-authored several of the papers we discussed here and who have always been very generous by making their work available
to others.  Geoff Clayton and David Silva are also working with us.  This
work is partially supported through the National Science Foundation  (AST-0604569).

\end{document}